\documentclass[twocolumn]{aastex62}

\usepackage{amsmath,amsfonts,graphicx,amssymb}
\usepackage{listings,hyperref,rotating,float}
\usepackage{xcolor, textcomp}
\usepackage{natbib}
\usepackage[nottoc]{tocbibind}
\usepackage{enumitem}
\usepackage[nottoc]{tocbibind}
\usepackage{rotating}
\usepackage{savesym}
\savesymbol{tablenum}
\usepackage{siunitx}
\restoresymbol{SIX}{tablenum}

\definecolor{mygreen}{RGB}{28, 172, 0}
\definecolor{mylilas}{RGB}{170, 55, 241}

\renewcommand{\thefootnote}{\alph{footnote}}

\numberwithin{equation}{section}

\newcommand\T{\rule{0pt}{2.8ex}}       % Top strut
\newcommand\B{\rule[-1.4ex]{0pt}{0pt}}       % Bottom strut
\renewcommand{\bold}[1]{\normalfont{#1}}

\let\oldFootnote\footnote
\newcommand\nextToken\relax

\renewcommand\footnote[1]{%
    \oldFootnote{#1}\futurelet\nextToken\isFootnote}

\newcommand\isFootnote{%
    \ifx\footnote\nextToken\textsuperscript{,}\fi}

\received{}
\revised{}
\accepted{}
\shorttitle{How many regions are necessary?}
\shortauthors{Whitbread et al.}

\begin{document}

\title{How many active regions are necessary to predict the solar dipole moment?}

%% The \author command is the same as before except it now takes an optional
%% arguement which is the 16 digit ORCID. The syntax is:
%% \author[xxxx-xxxx-xxxx-xxxx]{Author Name}
%%
%% Use \affiliation for affiliation information. Please use multiple
%% \affiliation calls for to document more than one affiliation.
%%
%% The new \altaffiliation can be used to indicate some secondary
%% information such as fellowships. NOTE that if an \altaffiliation
%% command is used it must come BEFORE the \affiliation call, right
%% after the \author command.
%%
%% Use \email to set provide email addresses. Each \email will appear
%% on its own line so you can put multiple email address in one \email
%% call. 
%%
%% A new \correspondingauthor command is available in V6.1 to identify the corresponding author of the manuscript. It is the author's responsibility to make sure this name is also in the author list.
%%
%% If done correctly the peer review system will be able to
%% automatically put the author and affiliation information from the manuscript
%% and save the corresponding author the trouble of entering it by hand.

\correspondingauthor{T. Whitbread}
\email{tim.j.whitbread@durham.ac.uk}

\author[0000-0001-9291-9011]{T. Whitbread}
\affiliation{Department of Mathematical Sciences, Durham University, Durham, DH1 3LE, UK}

\author[0000-0002-2728-4053]{A.~R. Yeates}
\affiliation{Department of Mathematical Sciences, Durham University, Durham, DH1 3LE, UK}

\author[0000-0002-4716-0840]{A. Mu\~noz-Jaramillo}
\affiliation{Southwest Research Institute, 1050 Walnut St. \#300, Boulder, CO 80302, USA}
\affiliation{National Solar Observatory, 3665 Discovery Drive, Boulder, CO 80303, USA}
\affiliation{High Altitude Observatory, National Center for Atmospheric Research, 3080 Center Green, Boulder, CO 80301, USA}

\begin{abstract}
  \noindent We test recent claims that the polar field at the end of Cycle 23 was weakened by a small number of large, abnormally oriented regions, and investigate what this means for solar cycle prediction. We isolate the contribution of individual regions from magnetograms for Cycles 21, 22 and 23 using a 2D surface flux transport model, and find that although the top $\sim$10\% of contributors tend to define sudden large variations in the axial dipole moment, the cumulative contribution of many weaker regions cannot be ignored. In order to recreate the axial dipole moment to a reasonable degree, many more regions are required in Cycle 23 than in Cycles 21 and 22 when ordered by contribution. We suggest that the negative contribution of the most significant regions of Cycle 23 could indeed be a cause of the weak polar field at the following cycle minimum and the low-amplitude Cycle 24. We also examine the relationship between a region's axial dipole moment contribution and its emergence latitude, flux, and initial axial dipole moment. We find that once the initial dipole moment of a given region has been measured, we can predict the long-term dipole moment contribution using emergence latitude alone.
\end{abstract}


\section{Introduction} \label{intro}

There is a strong correlation between the strength of the Sun's polar magnetic field at solar cycle minimum and the strength of the following cycle \citep[e.g.,][]{schatten78,andres13}. This means that it is possible to perform earlier solar cycle predictions by forecasting the evolution of the polar fields. Common methods for simulating the evolution of the radial magnetic field at the surface include using dynamo models \citep[for a review, see][]{charreview}, but surface flux transport (SFT) models \citep{wangetal89,baumann04,sheeleyreview, jiang10,mackayreview,hathuptonpredict,hathupton16}, introduced in the 1960s \citep{babcock,leighton64}, have risen in popularity over the last decade due to their relative simplicity and accuracy.

Surface flux transport models describe the evolution of magnetic regions on the solar surface, which appear due to the rise of buoyant flux tubes \citep{fan09}. Generally they emerge with a leading polarity and an opposing trailing polarity with respect to the east-west direction, and so are known as bipolar magnetic regions (BMRs). There is hemispheric asymmetry in the leading polarities, which are generally the same across a hemisphere, according to Hale's polarity law \citep{hale}. Helical convective motions in the solar interior impart a tilt to each BMR with respect to the east-west line (the line that connects the centres of the opposing polarities), with the leading polarity located closer to the equator. The effect is stronger at higher latitudes according to Joy's law \citep{joyslaw}, and a sinusoidal fit for the relationship between tilt angle $\alpha$ and latitude $\lambda$ is $\alpha = 32.1\sin\lambda$ \citep{stenflo12}, although it should be noted that there is significant variation between different regions. These deviations from Joy's Law could be the key characteristics in determining polar field strength at cycle minimum, as discussed below.

After emergence, the magnetic flux diffuses across the surface by being pushed to the edges of convection cells \citep{leighton64}, is advected poleward by meridional circulation, and sheared by differential rotation.  Due to the combined effects of Hale's and Joy's laws, the net result of this process is the cancellation of leading polarity flux across the equator and the accumulation of trailing polarity flux at the poles. This cancels the polar flux of the previous cycle and builds up new polar flux of the opposite polarity. It is this built-up polar field which provides an early insight into the amplitude of the following cycle.

Of particular interest is the unusually weak polar field (and equivalently weak axial dipole moment) at the end of Cycle 23 \citep{andres12}, which in turn is believed to be responsible for the low amplitude of Cycle 24. \citet{jiang15} used the BMR data of \citet{liulrich2012} to investigate the effect of tilt angle on axial dipole moment contribution $D$, using an empirical relation involving tilt angle, latitude and area \citep{jiang14}:
\begin{equation}\label{axdeqn}
  D \propto A^{\frac32}\sin\alpha\,\mbox{exp}\left(-\frac{\lambda^2}{110}\right) ,
\end{equation}
where $A$ is the area, $\alpha$ is the tilt angle, and $\lambda$ is the emergence latitude of each region. They found that axial dipole moment contributions from observed tilt angles in Cycle 23 follow those obtained by assuming Joy's Law at latitudes above $\pm 10$\textdegree{}. Nearer the equator, the regions with observed tilt angles contribute substantially less than would be expected from Joy's Law, contrary to the behaviour of Cycles 21 and 22, which follow the Joy's Law contributions more closely at all latitudes. This led to the suggestion that a single large anti-Hale or anti-Joy region emerging at a low latitude, or across the equator \citep{cameron13,cameron14}, has the ability to significantly alter the dipole moment, and this could have been the catalyst behind the weak polar field at the end of Cycle 23. Therefore the stochasticity behind the properties of emerging regions provides a problem for those attempting to predict the amplitude of future cycles, especially given that the magnetic flux in a single large active region is similar to the total polar flux \citep{wangsh91}. With this in mind, it may not be possible to make reliable predictions until the end of the cycle, unless random fluctuations of active region properties are taken into account. Indeed, \citet{nagy17} recently demonstrated in a 2$\times$2D dynamo model that large `rogue' regions can drastically affect the evolution of future solar cycles and introduce hemispheric asymmetries. Such large regions emerging during the early phases of a cycle can even affect the amplitude and duration of the same cycle. In this particular dynamo model, the effect of a single region can persist for multiple cycles. \citet{nagy17} found that the effect of a region in their model is dependent on its axial dipole moment at time of emergence, which is in turn approximated by Equation \ref{axdeqn}. So bipolar regions near the equator, and/or with large tilt angle, are are particularly strong contributors, although significant effects were found for regions even up to $\pm$20\textdegree{} latitude.

In this paper we investigate these claims further by simulating the evolution of real active regions from Cycles 21, 22 and 23 using a 2D SFT model\footnote{https://github.com/antyeates1983/sft\_data} with an automated region identification and assimilation process \citep{yeates15}. This allows us to identify particular observed properties which could have defined the contribution of each region to the axial dipole moment. In this paper, the emerging regions are determined from NSO line-of-sight magnetograms. In Section \ref{sect2} we discuss the extraction of regions and their properties in more detail. In Section \ref{regnum} we show how assimilating different numbers of regions based on both dipole moment contribution and flux can alter the end-of-cycle axial dipole moment. In Section \ref{sect4} we investigate in more detail how the properties of the regions determine their dipole contributions, before concluding in Section \ref{conclusions}.

\section{Determination of active region properties}\label{sect2}

We will investigate the distribution of various magnetic region properties, namely latitude, magnetic flux, and initial and final axial dipole moment. For each Carrington rotation in a cycle, the regions and their properties are extracted from NSO Kitt Peak or SOLIS synoptic magnetograms\footnote{http://solis.nso.edu/0/vsm/vsm\_maps.php} with resolutions of 180 pixels equally-spaced in sine-latitude and 360 pixels equally-spaced in longitude, and the overall photospheric evolution is simulated using the 2D SFT model described in \citet{yeates15}. The radial component of the magnetic field in 2D, $B\left(\theta ,\phi ,t\right)$, evolves according to:
\begin{align}\label{sfteqn}
  \frac{\partial{B}}{\partial{t}} =& - \omega \left(\theta\right)\frac{\partial{B}}{\partial{\phi}} - \frac{1}{R_{\odot}\sin\theta}\frac{\partial}{\partial{\theta}}\Big(v\left(\theta\right)\sin\theta\,B\Big) \nonumber\\
  & + \frac{\eta}{R_{\odot}^2}\left[\frac{1}{\sin\theta}\frac{\partial}{\partial{\theta}}\left(\sin\theta\frac{\partial{B}}{\partial{\theta}}\right) + \frac{1}{\sin^2\theta}\frac{\partial ^2{B}}{\partial{\phi ^2}}\right] \nonumber\\
  & - \frac{1}{\tau}B + S(\theta,\phi,t) ,
\end{align}
where $R_{\odot}$ is the solar radius, $\omega\left(\theta\right)$ represents differential rotation, $\eta$ is turbulent diffusivity, representing the diffusive effect of granular convective motions, $\tau$ is an exponential decay term added by \citet{schrijver02} to improve regular polar field reversal, and $S(\theta,\phi,t)$ is a source term for newly emerging magnetic regions. The profile $v\left(\theta\right)$ describes poleward meridional flow, which we define using the following functional form: 
\begin{equation}\label{meridfloweqn}
  v\left(\theta\right) = -v_0\sin^p\theta\cos\theta ,
\end{equation}
where $p$ determines the latitude of peak velocity and low-latitude gradient. For the initial condition, we use the profile of \citet{svalgaard78}:
\begin{equation}\label{initeqn}
  B\left(\theta ,0\right) = B_0\,\lvert\cos\theta\rvert ^7 \cos\theta ,
\end{equation}
where $B_0$ is the initial field strength. The new magnetic regions comprising the source term are determined from synoptic magnetograms and each region is assimilated on the day when its centroid crosses the central meridian. The assimilation algorithm is described fully in the Appendix of \citet{yeates15}. Briefly, the synoptic magnetograms are corrected for flux imbalance, then their absolute value is smoothed with a Gaussian filter (standard deviation $\sigma = 3$), so as to merge positive and negative polarities. Each region is then determined by a connected group of pixels above the threshold $B_{\rm par}$, which is set to the same value $B_{\rm par} = 39.8$\,G as found in \citet{me17}. These pixels (from the original unsmoothed synoptic map) are then inserted into the simulation, replacing any pre-existing $B_r$ in that region. The flux is corrected so as to preserve the pre-existing net flux in that region of the simulation.

The evolution equations for the vector potential are solved in the Carrington frame using a finite-difference method on a grid with a resolution of 180 equally-spaced pixels in both sine-latitude and longitude. The model is fully automated and is constructed such that new regions replace pre-existing ones, rather than being superimposed on them. In some cases, very strong regions can reappear in the magnetogram of the following Carrington rotation. Because of complex flux emergence and cancellation processes that occur between the multiple observations of the same region, it is not trivial to automatically define whether an active region is new or a repeat in the model, so we class these repeats as new regions altogether, and the replacement technique ensures that the axial dipole moment contribution from a returning region is not counted twice. This method ensures that the repeated regions do not affect our conclusions.

\begin{figure}
  \resizebox{\hsize}{!}{\includegraphics{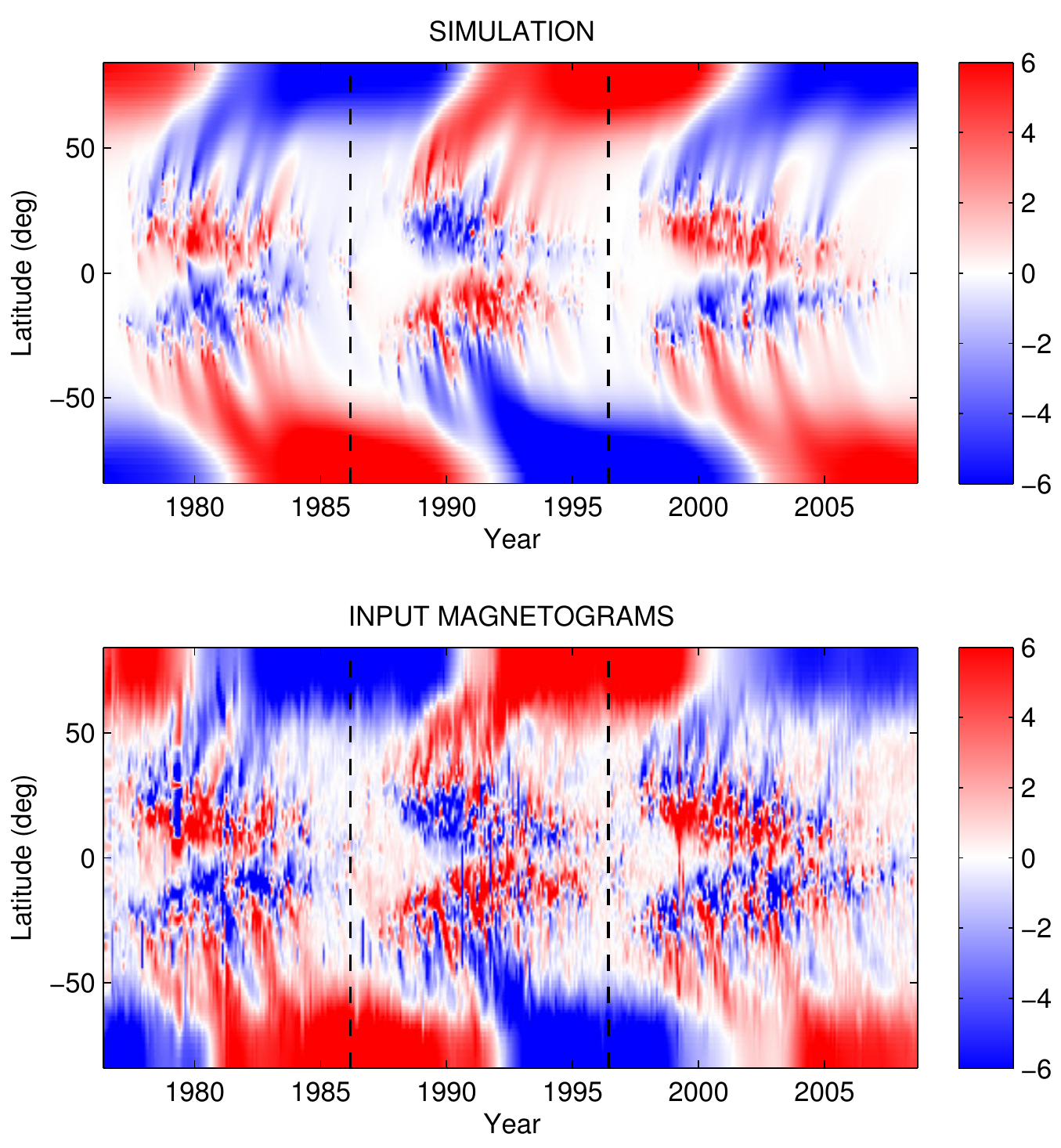}}
  \caption{Top: Optimal butterfly diagram for Cycle 21 through to Cycle 23, simulated using the parameters from Table \ref{table1}. Bottom: `Ground truth' data for the same period. Vertical dashed lines indicate start/end points of cycles as used in this paper.}\label{optbfly}
\end{figure}

All simulations are performed using optimal values for diffusivity, meridional flow, initial field strength, exponential decay and assimilation threshold, obtained using the genetic algorithm \texttt{PIKAIA}\footnote{http://www.hao.ucar.edu/modeling/pikaia/pikaia.php}\footnote{http://www.hao.ucar.edu/Public/about/Staff/travis/mpikaia/} \citep{pikaia,mpikaia,lemerle}, as described in \citet{me17}. The present optimum values are shown in Table \ref{table1}, with associated `acceptable ranges' below each entry. Note that we keep these parameters fixed across the three cycles, and that $B_0$ is the initial field strength at the start of Cycle 21; each other cycle immediately follows on from the final state of the preceding cycle. `Optimal' in this sense refers to the ability to best match the simulated and observed butterfly diagrams, and the optimal butterfly diagram for Cycles 21 to 23 is shown in the top panel of Figure \ref{optbfly}. The bottom panel shows the observed butterfly diagram obtained from full-disk images from US National Solar Observatory, Kitt Peak, which underwent a polar field correction procedure described by \citet{petrie12}. All conclusions made in this paper are with respect to these optimal parameter values. For differential rotation, the parametrization of \citet{snodgrass90} is used:
\begin{equation}
  \omega \left(\theta\right) = 0.521 - 2.396\cos^2\theta - 1.787\cos^4\theta\,\mbox{deg}\,\mbox{day}^{-1} .
\end{equation}
We also include an exponential decay term of the form $-\frac{1}{\tau}B$. \citet{baumann06} offered a physical explanation for the extra term: it is the effect of inward radial diffusion of flux into the convection zone, which is not directly accounted for in the SFT model. In Appendix \ref{appendix1} we present the case without decay and show that similar conclusions hold in both regimes.

\begin{table*}
  \caption{Optimal parameter set for the simulation shown in Figure \ref{optbfly}. Upper and lower bounds for acceptable parameter ranges are given in square brackets below each entry, although here we use the optimum values themselves for all simulations.}
  \label{table1}
  \centering
  \begin{tabular}{c c c c c c c c}
    \hline
    $\eta$ & $v_0$ & $p$ & $\tau$ & $B_0$ \T\\
    (km$^2$\,s$^{-1}$) & (m\,s$^{-1}$) & & (yr) & (G) \B\\
    \hline
    466.8 & 9.2 & 2.33 & 10.1 & 6.7 \T\\
    $\left[325.7,747.3\right]$ & $\left[5.6,11.9\right]$ & $\left[1.12,3.95\right]$ & $\left[3.6,31.9\right]$ & $\left[0.0,15.0\right]$ \B\\
    \hline
  \end{tabular}
\end{table*}

The axial dipole moment of region $i$ is given by:
\begin{equation}\label{axdieqn}
  D^{\left(i\right)}\left(t\right) = \frac32 \int_0^{\pi}\int_0^{2\pi} B^{\left(i\right)}\left(\theta,\phi,t\right)\cos\theta \sin\theta\,d\phi\,d\theta ,
\end{equation}
where $B^{\left(i\right)}\left(\theta,\phi,t\right)$ is the evolving magnetic field of the individual region $i$, computed after its initial insertion by solving Equation \ref{sfteqn} with no other field present. Isolating the evolution of a single region like this is meaningful because Equations \ref{sfteqn} and \ref{axdieqn} are approximately linear, so that the contributions $D^{\left(i\right)}\left(t\right)$ may be added together to give the overall dipole moment $D_{\rm tot}\left(t\right)$. The linearity is only approximate because our newly inserted regions replace pre-existing flux, and strong returning regions from the previous rotation are treated as new regions, as discussed above. Nevertheless, the evolution of the strongest of a set of repeated regions is a good approximation to the combined evolution including replacements, and it is therefore useful to isolate them.

To assess the contribution of each region to the overall evolution of the dipole moment, we will also use the relative axial dipole moment $D_{\rm rel}$, which is defined as:
\begin{equation}
  D^{\left(i\right)}_{\rm rel}\left(t\right) = \frac{D^{\left(i\right)}\left(t\right)}{D_{\rm tot}\left(t_{\rm end}\right) - D_{\rm tot}\left(t_{\rm start}\right)} ,
\end{equation}
\bold{for region $i$, where $D_{\rm tot}\left(t\right)$ is the dipole moment of the full simulation with all regions included, and $D^{\left(i\right)}\left(t\right)$ is the dipole moment contribution of a single active region as calculated in Equation \ref{axdieqn}.} The times $t_{\rm start}$ and $t_{\rm end}$ are the start and end of each cycle respectively, so that $D_{\rm rel}^{\left(i\right)}$ represents the contribution from region $i$ to the overall change in dipole moment during the cycle. The start and end times are set to: $t_{\rm start} =$ 1976 May 1 and $t_{\rm end} =$ 1986 March 10 for Cycle 21, $t_{\rm start} =$ 1986 March 10 and $t_{\rm end} =$ 1996 June 1 for Cycle 22, and $t_{\rm start} =$ 1996 June 1 and $t_{\rm end} =$ 2008 August 3 for Cycle 23.  The final relative axial dipole moment $D_{\rm rel}^{\left(i\right)}\left(t_{\rm end}\right)$ then reflects the proportional contribution of region $i$ to the end-of-cycle axial dipole moment. A positive $D_{\rm rel}\left(t_{\rm end}\right)$ corresponds to a strengthening of the axial dipole moment at the end of the cycle, whilst a negative $D_{\rm rel}\left(t_{\rm end}\right)$ corresponds to a weakening.

\begin{figure}
 \resizebox{\hsize}{!}{\includegraphics{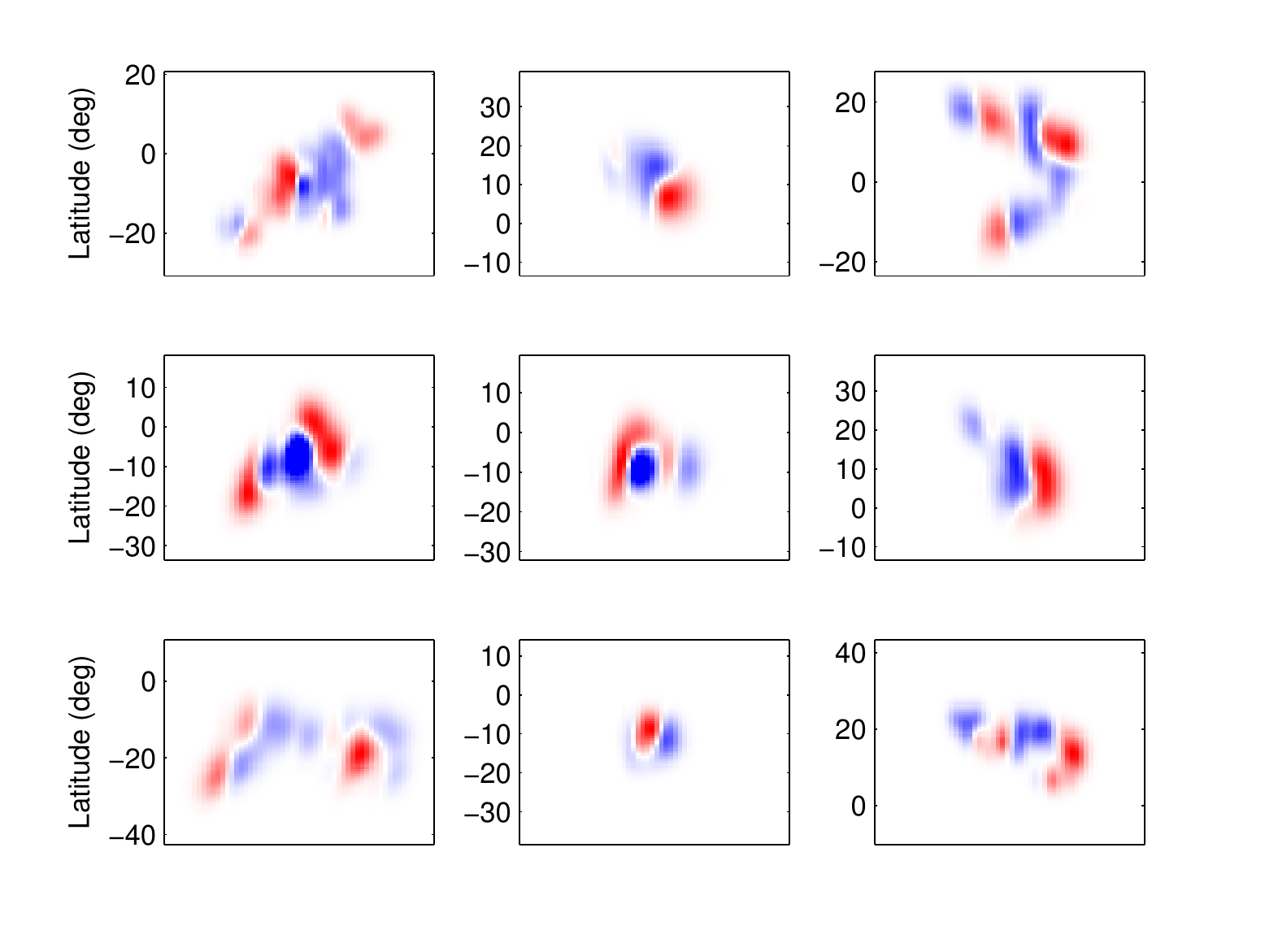}}
  \caption{Nine most significant contributing regions from Cycle 23, as measured by $D_{\rm rel}\left(t_{\rm end}\right)$. The panels are equal in size and centred around each region. Each image is saturated individually.}\label{9regsc23}
\end{figure}

Note that most SFT simulations, including \citet{jiang15}, assume that all regions are BMRs with a simple bipolar structure. However in our 2D model this is not always the case. The model inserts the observed shapes of active regions, meaning that complex multipolar configurations are often assimilated. Figure \ref{9regsc23} shows the configurations of the top nine largest contributors from Cycle 23, as measured by $D_{\rm rel}\left(t_{\rm end}\right)$. Among these are two regions that share similar features (left and centre panels of the middle row), and are likely to have been the same region appearing in two consecutive rotations, having undergone some sort of interaction in the interim. Whilst some regions are clearly bipolar, some are less clear and are harder to separate into BMRs. Because of this, a `tilt angle' is no longer a sensible measure, and so instead we use the initial (relative) axial dipole moment which still takes into account orientation and polarity. Similarly, we also do not consider polarity separation distance. Here the initial axial dipole moment of an active region is measured at the time of assimilation, that is, on the day it crosses the central meridian.

For the optimal threshold $B_{\rm par}$, we tend to extract fewer regions per cycle than other studies, because the model can consider a cluster of active regions to be one single large region. Despite this, the insertion of realistic configurations of active regions combined with the optimization procedure means that the evolution of the observed axial dipole moment $D_{\rm tot}$ is well reproduced by the simulation, even though the axial dipole moment is not considered directly in the fitness function (unlike \citet{lemerle}). We will also continue to use the term `regions' to describe both individual and clusters of regions.

\section{How many regions are required?}\label{regnum}

\begin{figure*}
  \resizebox{\hsize}{!}{\includegraphics{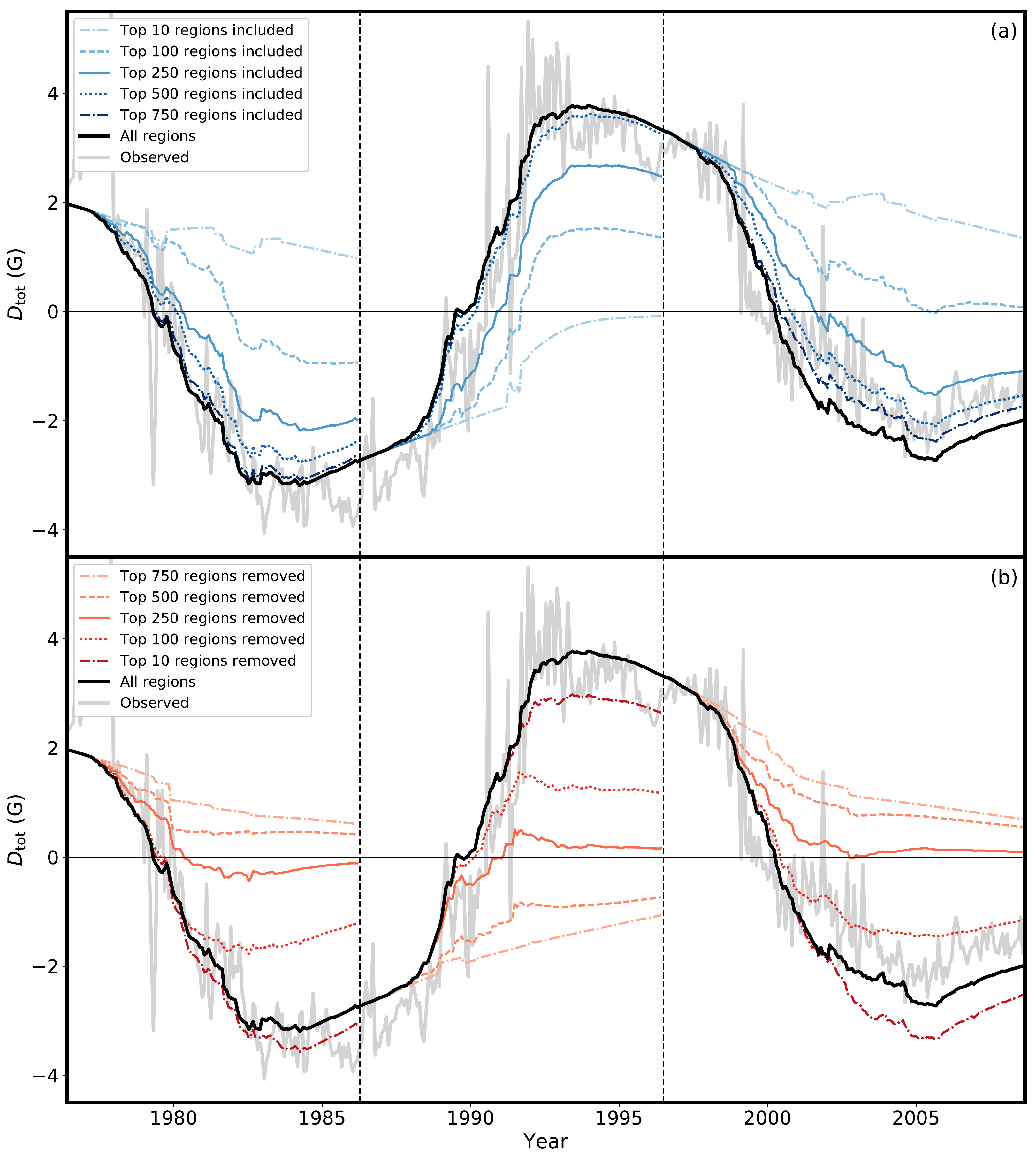}}
  \caption{Evolution of the axial dipole moment for Cycles 21 to 23. Each profile is obtained by: (a) only using a certain number of the biggest contributors to the axial dipole moment, or (b) removing the biggest contributors to the axial dipole moment. Colour intensity is indicative of the number of regions used in each simulation, as shown in the legend. The light grey curve shows the observed axial dipole moment. Vertical dashed lines indicate start/end points of cycles as used in this paper.}
  \label{5profiles}
\end{figure*}

Initially we consider the effect on the overall axial dipole moment of including the largest dipole moment contributions only, to assess how many regions are needed to replicate the original axial dipole moment. Regions are listed in order of absolute $D_{\rm rel}\left(t_{\rm end}\right)$ and only those above a certain threshold are assimilated. This routine is performed at five thresholds so that the top 10, 100, 250, 500 and 750 regions are included over five separate runs in each cycle, and the resulting profiles are shown in Figure \ref{5profiles}(a). These are superimposed on the observed axial dipole moment (light grey).

The left-hand section of Figure \ref{5profiles}(a) shows the effect of keeping the largest contributions to the axial dipole moment from the simulation of Cycle 21. Incorporating the largest 750 contributors of the 844 regions makes only a little difference (a decrease of 1.6\%), but using 500 regions corresponds to a reduction of 7\% of the axial dipole moment.

The middle section of Figure \ref{5profiles}(a) shows the effect of including the largest contributions to the axial dipole moment from the simulation of Cycle 22. As few as 500 of the 846 regions can be used with a shortfall of just 1.3\%, and using 750 regions makes little difference to the evolution of the axial dipole moment. If we assimilate the top ten contributors of Cycle 22, polar field reversal is almost achieved.

The right-hand section of Figure \ref{5profiles}(a) shows the same profiles as the left and middle sections but for Cycle 23. Even when the largest 750 contributors of the 951 regions are assimilated, there is a more significant discrepancy (a decrease of 4.7\%) between the resulting axial dipole moment and $D_{\rm tot}$ than in the previous two cycles. We will show later that this is because most of the large contributors in Cycle 23 act to weaken the overall dipole moment (opposite to the majority pattern). The cumulative contribution of many weaker regions is therefore needed to recover its final strength. So although a small number of regions have a disproportionate effect, the cumulative contribution of the many regions with weaker dipole moment cannot be ignored, owing to their common sign.

In each cycle we see that the top $\sim$10\% of contributors (that is, about 100 of them) determine the rapid short-term changes in the axial dipole moment. Here we see the deficit in Cycle 23; even when the top 100 contributors are included the polar field is still unable to reverse. If we remove the top 10 strongest regions from the simulation instead of keeping them (Figure \ref{5profiles}(b)), we discover that the amplitude of the final axial dipole moment is overestimated in Cycles 21 and 23, and underestimated in Cycle 22. This demonstrates the impact of the strongest regions from the three cycles, and that the polar field at the end of Cycle 23 could have been stronger had the strongest few regions emerged with different properties or not emerged at all. If the top 100 strongest regions are removed from Cycle 23, the axial dipole moment is better represented than in the equivalent cases for Cycles 21 and 22, presumably because the proportion of regions with negative dipole contribution is greater in Cycle 23.

%It is imperative to note that there are more total regions involved in our simulations of Cycle 23 than in Cycles 21 and 22, because it is a weaker and therefore longer cycle. The consequence of this is that the same number of regions in Cycle 23 represents a smaller proportion of the total number of regions compared to the other two cycles, and so naturally we might expect the axial dipole moment to be weaker when using, say, 750 regions in Cycle 23. However, when we add more regions to Cycle 23 to balance the proportion with other cycles, there is still a larger difference between the profile with all regions and the profile with some regions removed. We conclude that ultimately this pattern comes down to the polarity distribution of regions with a small contribution to the axial dipole moment, and that the smallest 100 contributors of Cycle 23 must have predominantly positive $D_{\rm rel}\left(t_{\rm end}\right)$.%

\subsection{What are the implications for making predictions?}

\begin{figure}
  \resizebox{\hsize}{!}{\includegraphics{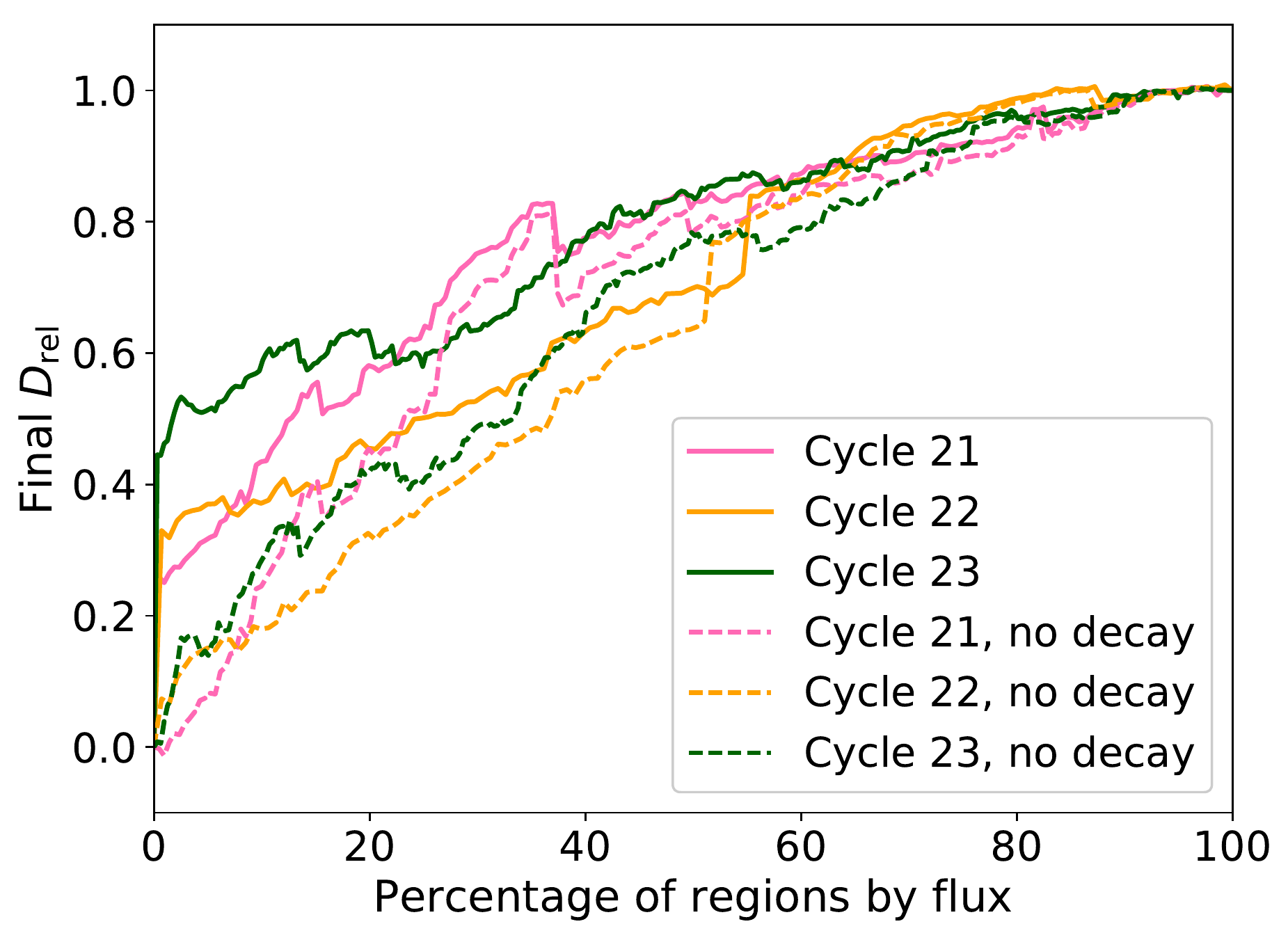}}
  \caption{Final $D_{\rm rel}$ against percentage of regions included for Cycles 21 (pink), 22 (yellow) and 23 (dark green). Solid lines are the cases with exponential decay, and dashed lines are the cases where the decay term has been removed. Regions are ordered by flux and the top $x\%$ of the strongest regions are incorporated.}
  \label{reg_percent}
\end{figure}

Up to this point regions have been ordered by $D_{\rm rel}\left(t_{\rm end}\right)$. Unfortunately, calculating this at time of emergence requires us to know the subsequent behaviour of all other regions during the rest of the cycle. So we now examine the consequences of ordering and including regions based on absolute flux, which is a quantity readily measured at time of emergence. The solid lines in Figure \ref{reg_percent} display the change in $D_{\rm rel}\left(t_{\rm end}\right)$ as more active regions are included in the simulation, ordered by decreasing flux, for Cycles 21 (pink), 22 (yellow) and 23 (dark green).

There are multiple regions with large flux that contribute positively to the axial dipole moment during Cycle 21. Because of this, 80\% of $D_{\rm tot}\left(t_{\rm end}\right)$ is attained when less than 40\% of regions are considered (bearing in mind the threshold for the top 40\% is $\sim$\,4--\num{4.5e21}\,Mx depending on the cycle). There is then a sharp decrease when the two biggest contributions of $D_{\rm rel}\left(t_{\rm end}\right)$ are included, before the 80\% mark is reached again, corresponding to half the number of regions being used. Note that more than 25\% of $D_{\rm tot}\left(t_{\rm end}\right)$ is attained by using only a small percentage of the largest regions. This is a side-effect of the measure we use. For example, when decay is not present (see Figure \ref{5profiles_nodecay} in Appendix \ref{appendix1}) and 10 regions are included, the end-of-cycle dipole moment is far away from the original end-of-cycle dipole moment (thick black line), and the contribution is small (dashed profiles in Figure \ref{reg_percent}). However when we include decay (Figure \ref{5profiles}), these profiles both go closer to zero, thereby reducing the difference between the two end-of-cycle dipole moments and hence increasing the \textit{relative} dipole moment obtained by the 10 regions. This effect is even stronger for the other two cycles. Inclusion of decay does not affect the basic shape of each profile, it merely weakens the contribution from stronger regions. This can be seen by comparing the solid and dashed lines in Figure \ref{reg_percent}.

The $D_{\rm rel}\left(t_{\rm end}\right)$ of Cycle 22 rises at a steady rate as more regions are added, but there are two clear phases with a large jump in between. One can attribute this jump to the inclusion of the largest contributor of Cycle 22. Because of this significant addition to the dipole moment, using 55\% of regions is enough to ensure that 80\% of $D_{\rm tot}\left(t_{\rm end}\right)$ is reached.

The profile for Cycle 23 initially reaches almost 0.5\,$D_{\rm tot}\left(t_{\rm end}\right)$, presumably because the regions with strongest flux contribute positively to the dipole moment. There is then barely an increase in $D_{\rm rel}\left(t_{\rm end}\right)$ as another 30\% of the regions are included. This mimics the problem found in Figure \ref{5profiles}; Cycle 23 is largely dominated by negative $D_{\rm rel}\left(t_{\rm end}\right)$ active regions.

It may be noteworthy that when 60\% of the strongest regions are incorporated (i.e. regions with flux above about \num{2e21}\,Mx), the three cycles reach 80\% of the final $D_{\rm tot}$ and adding small regions bears minimal difference, regardless of cycle number. If 90\% of regions are used, corresponding to a threshold of approximately \num{5e20}\,Mx, all three cycles reach a similar relative level close to $D_{\rm tot}\left(t_{\rm end}\right)$.

\section{Distributions of active region properties}\label{sect4}

\subsection{Latitude, flux and initial dipole moment}\label{locsize}

\begin{figure*}
  \gridline{\fig{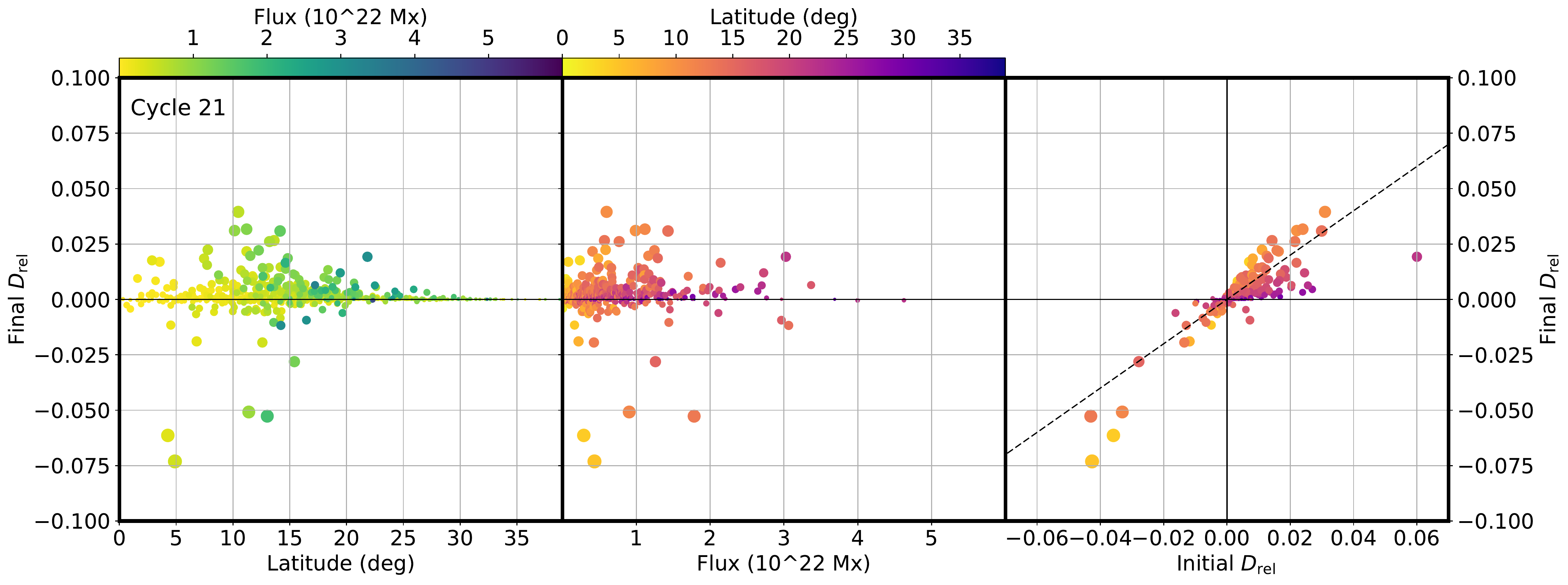}{\textwidth}{}}
  \gridline{\fig{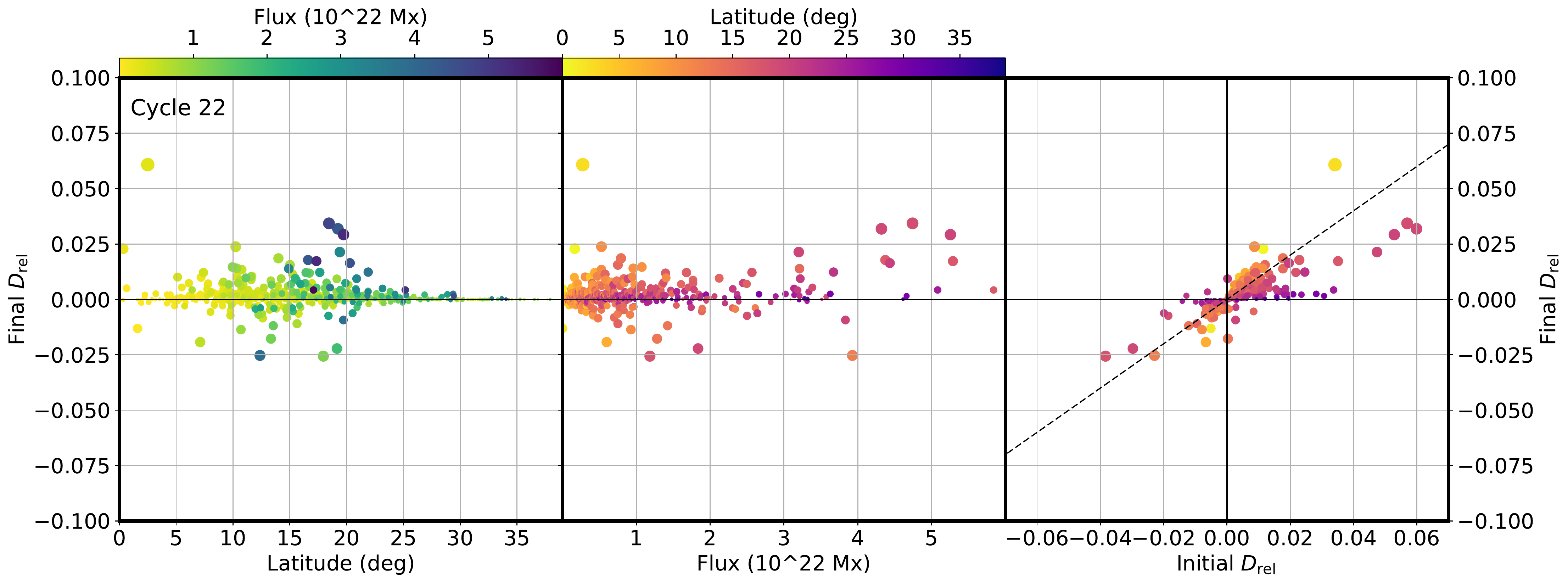}{\textwidth}{}}
  \gridline{\fig{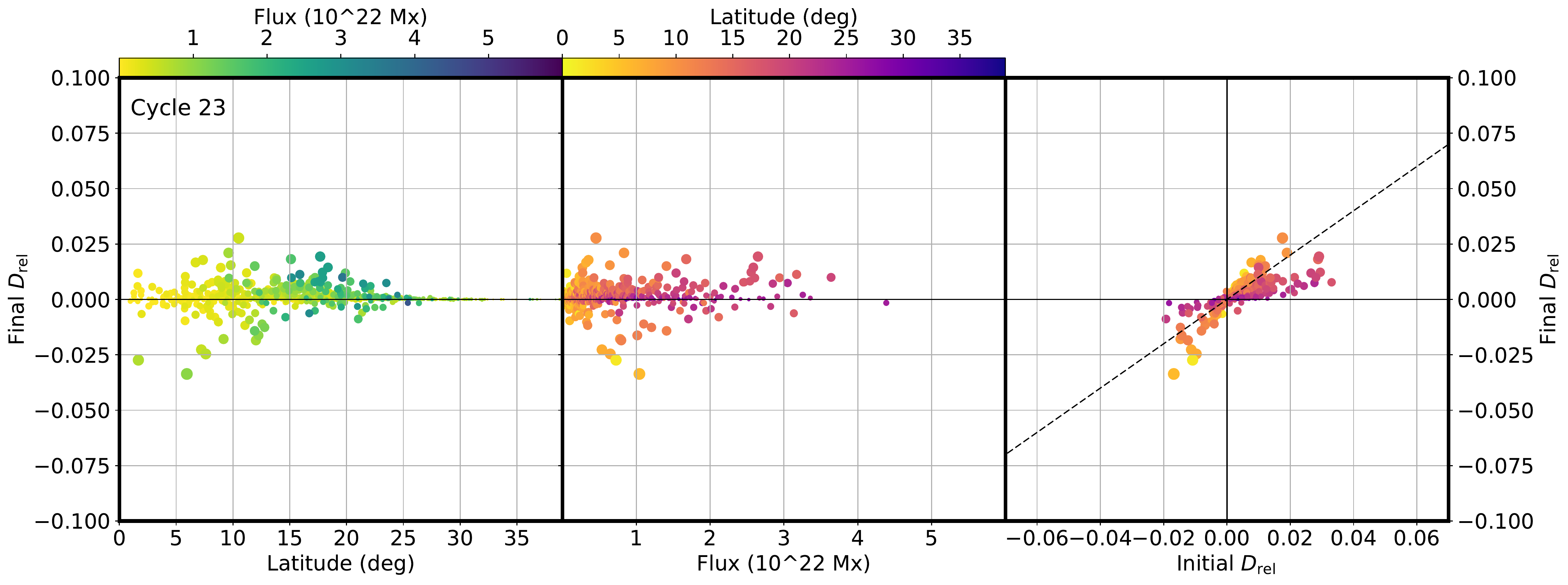}{\textwidth}{}}
  \caption{Final $D_{\rm rel}$ for each region against absolute latitude (left panels), flux (middle panels) and initial $D_{\rm rel}$ (right panels). Markers are sized by absolute final $D_{\rm rel}$, and coloured by flux (left panels) and absolute latitude (middle and right panels).}\label{allcycles_3panels}
\end{figure*}

We now turn to analyse the effects of emergence latitude, flux and initial $D_{\rm rel}$ on the axial dipole moment contribution $D_{\rm rel}\left(t_{\rm end}\right)$ of each region. The top panels of Figure \ref{allcycles_3panels} show the relationships between $D_{\rm rel}\left(t_{\rm end}\right)$ and these three quantities from left to right respectively for the regions from Cycle 21. We find that most significant contributors to the axial dipole moment emerge below $\pm 20$\textdegree{}, the very largest of which emerge below $\pm 10$\textdegree{}. We also find that these regions do not necessarily have strong levels of magnetic flux; very few of the biggest contributors are stronger than \num{1.5e22}\,Mx.

We discover that the relationship between initial and final $D_{\rm rel}$ is largely determined by the emergence latitude: regions emerging at mid-latitudes (dark purple) tend to contribute little to the final axial dipole moment, regardless of their initial values. Conversely, regions emerging at low latitudes (yellow and orange) can undergo an increase in axial dipole moment contribution as cross-equatorial flux cancellation occurs and flux is transported poleward by the meridional flow.

The central row of Figure \ref{allcycles_3panels} shows the same relationships as discussed above but for Cycle 22. The left and middle panels tell a different story to that of Cycle 21. There are fewer big contributions (i.e. contributions of more than 2.5\%) to the axial dipole moment, and the largest is a strengthening rather than a weakening as in Cycle 21. This explains why the axial dipole moment increased in amplitude during Cycle 22, and why polar field reversal is almost achieved with just ten regions in Figure \ref{5profiles}(a). This largest region is also the only significant contributor to lie below $\pm 10$\textdegree{}, although the others still emerge below $\pm 20$\textdegree{} as in Cycle 21. The most striking difference between the two cycles is the effect of strong-flux regions. In Cycle 22 some of the most significant contributions to the axial dipole moment come from regions with fluxes above \num{3e22}\,Mx, which is not the case in Cycle 21. The same latitudinal dependence of the initial to final $D_{\rm rel}$ relationship is found as in Cycle 21, supporting the idea that latitude of emergence plays an important role in determining whether a region will contribute significantly to the polar field.

The bottom three panels of Figure \ref{allcycles_3panels} show the same three distributions but for Cycle 23. We return to a similar regime to Cycle 21: of the most significant contributors, we observe more regions which weaken the axial dipole moment, and the biggest contributors have fluxes smaller than \num{2e22}\,Mx. Again, most of these regions emerge below $\pm 20$\textdegree{}. We find that the most significant regions in Cycle 23 induce a weakening of the overall axial dipole moment. These low-latitude regions could indeed be the cause of the weak polar field at the end of Cycle 23, and hence the low amplitude of Cycle 24, as suggested by \citet{jiang15}.

\begin{figure}
  \resizebox{\hsize}{!}{\includegraphics{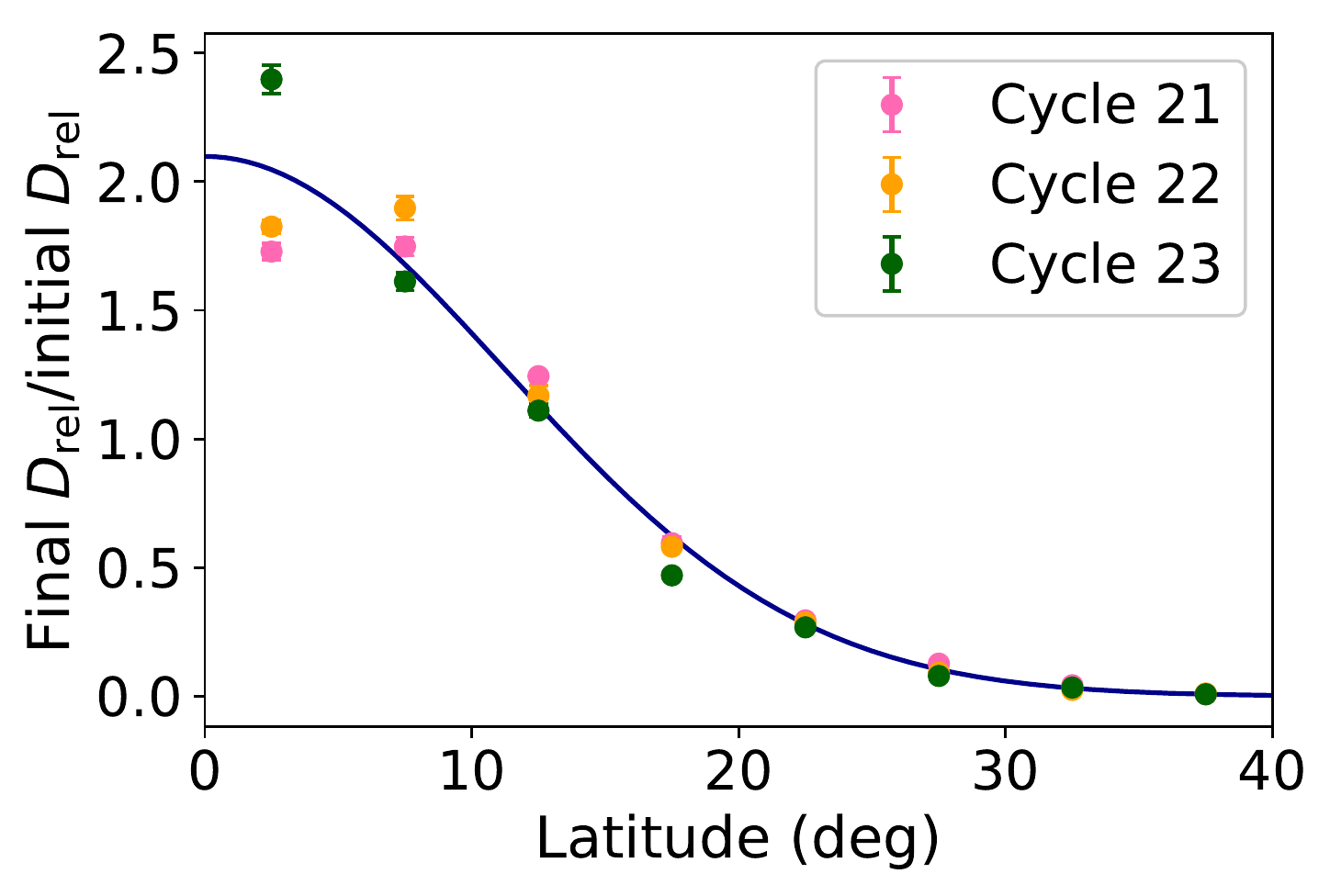}}
  \caption{Ratio between final $D_{\rm rel}$ and initial $D_{\rm rel}$ for 5\textdegree{} latitudinal bins for Cycles 21 (pink), 22 (yellow) and 23 (dark green). Error bars show standard error. Markers are plotted at the midpoint of each 5\textdegree{} bin. The dark blue curve is a Gaussian fit to the data.}
  \label{ratioscatter}
\end{figure}

The latitude-dependent relationship between initial and final $D_{\rm rel}$ still holds in Cycle 23. Separating the regions into bins of 5\textdegree{} and calculating the gradient of the lines in the right-hand panels of Figure \ref{allcycles_3panels} for each bin (see Figure \ref{ratioscatter}), we find that down to $\pm 20$\textdegree{} the relationship between initial and final $D_{\rm rel}$ is practically identical across the three cycles, and even down to $\pm 5$\textdegree{} the relationships over the three cycles are close. For the 0--5\textdegree{} bin, the gradient is much steeper for Cycle 23. However, this bin has relatively few points, and is least well fitted by a linear relationship between initial and final $D_{\rm rel}$. The standard errors for these fits are very small, indicating a strong relationship between the overall amplification in $D_{\rm rel}$ and the latitude of emergence. If we fit a Gaussian to the data (dark blue curve in Figure \ref{ratioscatter}), we find that the axial dipole moment contribution is proportional to $\mbox{exp}\left(-\frac{\lambda^2}{252}\right)$. This is similar to the relationship between latitude and axial dipole moment contribution given by \citet{jiang14} who also found a Gaussian latitudinal dependence in their model (Equation \ref{axdeqn}).

\subsection{Latitude and time}\label{loctime}

\begin{figure*}
  \resizebox{\hsize}{!}{\includegraphics{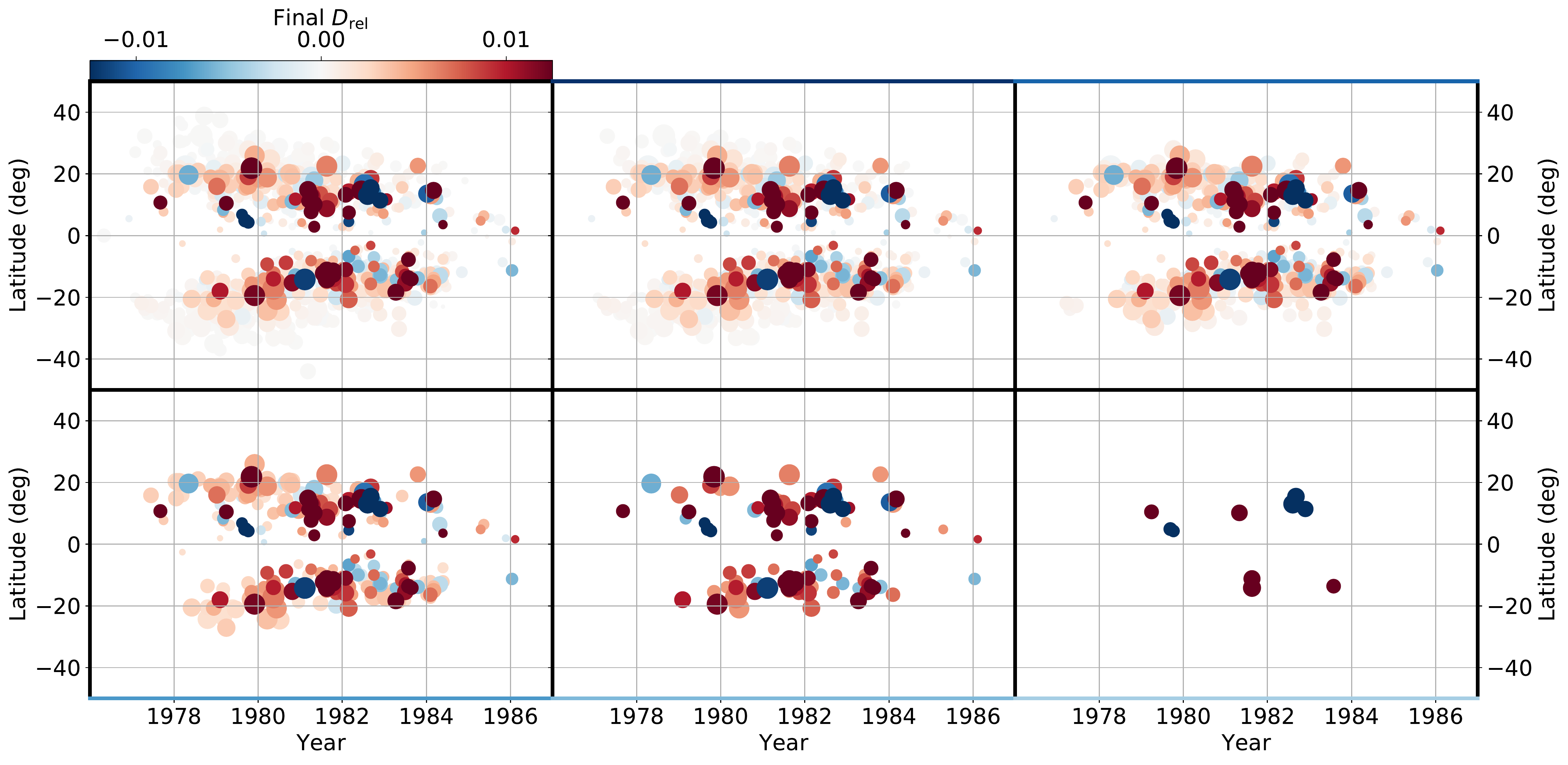}}
  \caption{Time-latitude distributions of regions from Cycle 21 used to obtain the profiles in the first section of Figure \ref{5profiles}(a) (profile colours match border colours). Markers are sized by flux and coloured by final $D_{\rm rel}$.}
  \label{6panelc21}
\end{figure*}

We now focus on the time-latitude distributions, i.e. `butterfly diagrams', of the active regions drawn from the assimilative 2D model. Figure \ref{6panelc21} shows the butterfly diagrams of Cycle 21 for the cases shown in the first section of Figure \ref{5profiles}(a), where border colours match profile colours. We find few strong regions that have emerged across the equator, suggesting that large contributors from Cycle 21 are likely to be because of orientation reasons rather than being cross-equatorial. There is a cluster of negatively contributing regions in the northern hemisphere around 1983 which is not followed by many significant regions during the remainder of the cycle; this cluster could be responsible for a lower axial dipole moment in Cycle 21 (compared to Cycle 22), and explains why the polar field fails to reverse when only 10 regions are used in Cycle 21, as seen in Figure \ref{5profiles}(a).

\begin{figure*}
  \resizebox{\hsize}{!}{\includegraphics{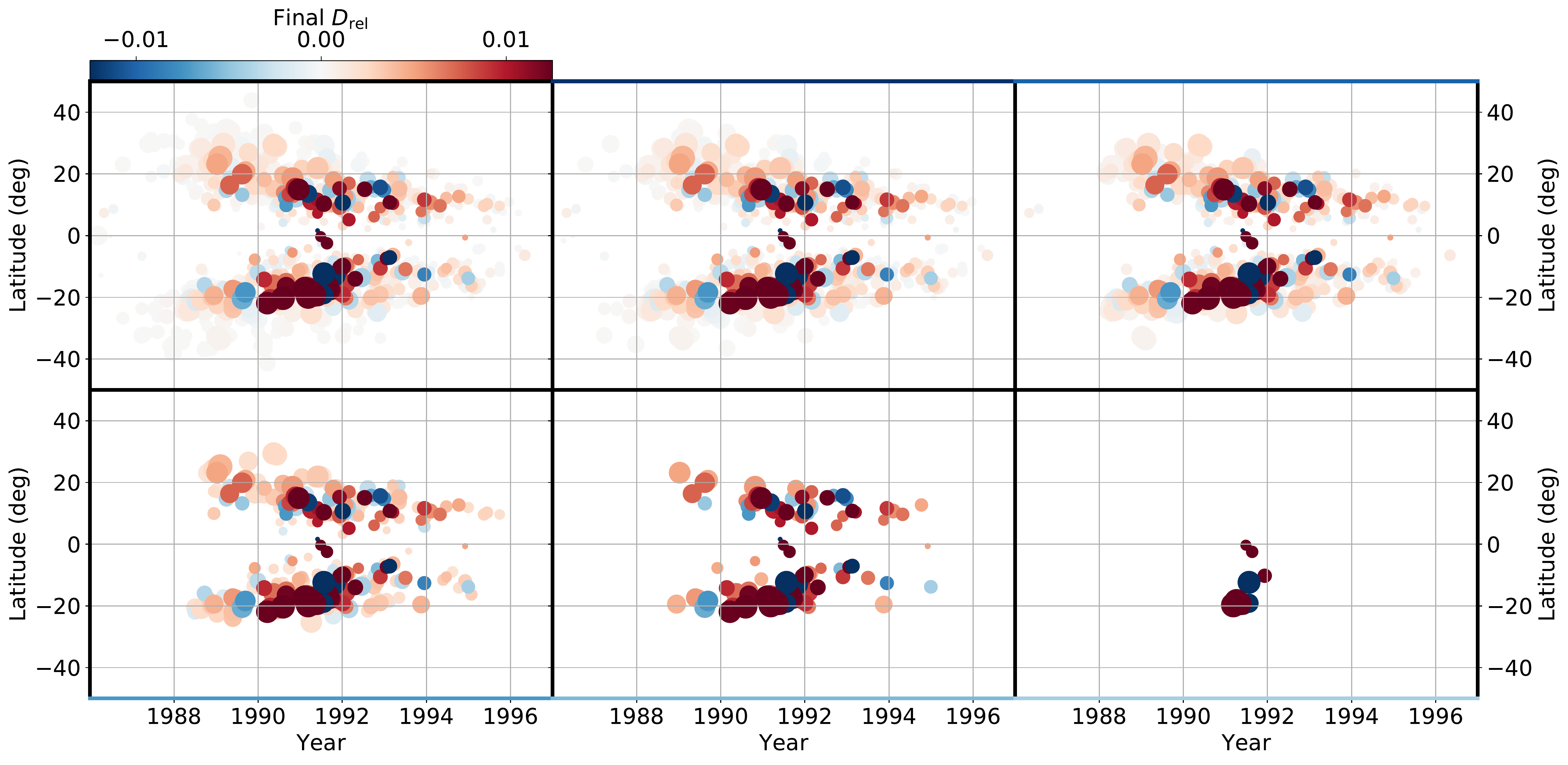}}
  \caption{Time-latitude distributions of regions from Cycle 22 used to obtain the profiles in the second section of Figure \ref{5profiles}(a) (profile colours match border colours). Markers are sized by flux and coloured by final $D_{\rm rel}$.}
  \label{6panelc22}
\end{figure*}

Figure \ref{6panelc22} shows the corresponding butterfly diagrams for Cycle 22. As inferred from Figure \ref{allcycles_3panels}, the majority of large contributions to the axial dipole moment in Cycle 22 enhance the dipole moment and are clustered around -20\textdegree{}. However, there are two large contributors at low latitudes, possibly cross-equatorial, which would support the claim of \citet{cameron13}: that regions emerging across the equator can significantly change the amount of net flux in each hemisphere, in turn weakening or strengthening the axial dipole moment, meaning future cycle predictions will be less reliable.

\begin{figure*}
  \resizebox{\hsize}{!}{\includegraphics{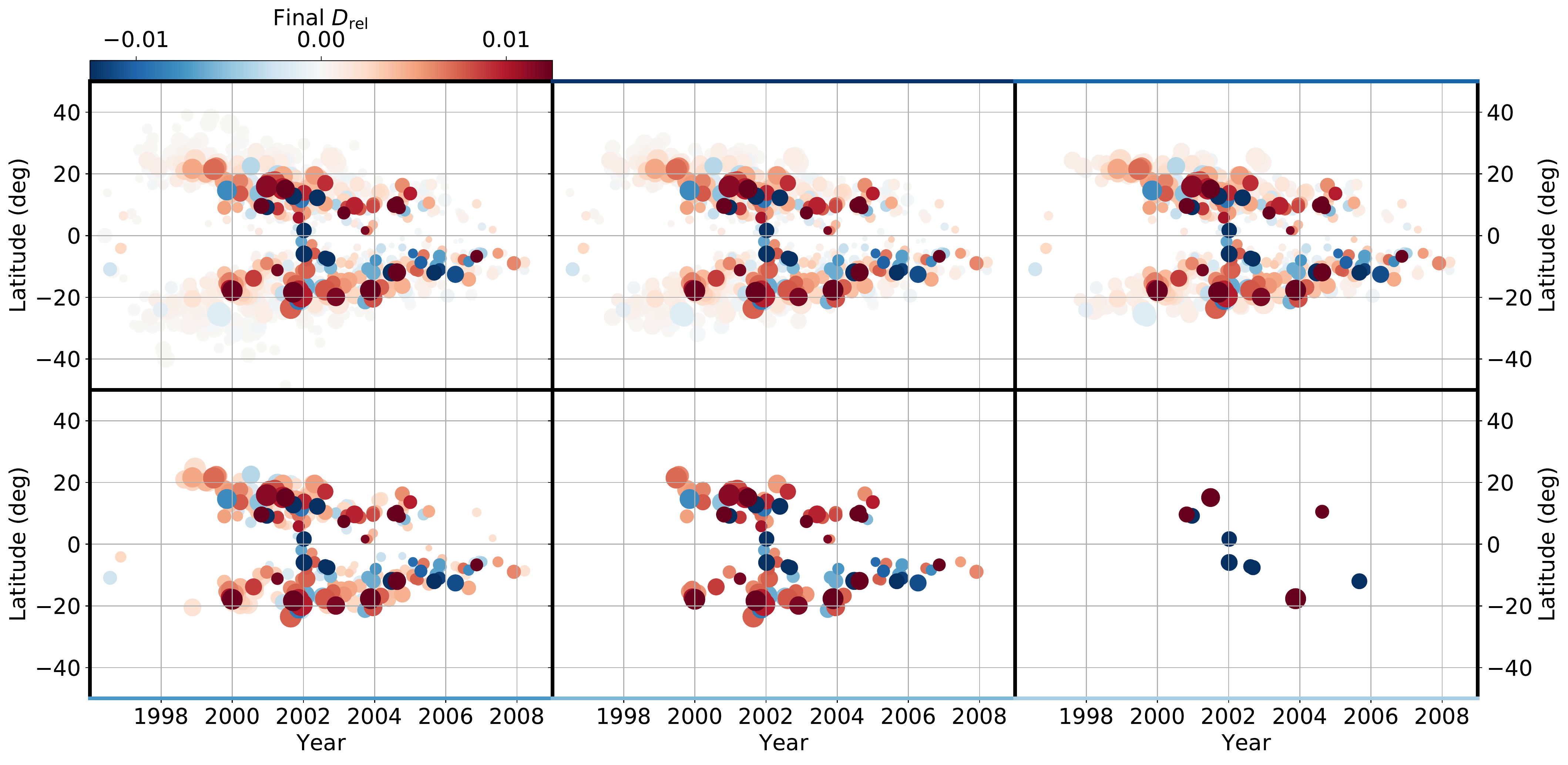}}
  \caption{Time-latitude distributions of regions from Cycle 23 used to obtain the profiles in the third section of Figure \ref{5profiles}(a) (profile colours match border colours). Markers are sized by flux and coloured by final $D_{\rm rel}$.}
  \label{6panelc23}
\end{figure*}

Figure \ref{6panelc23} shows the butterfly diagrams of Cycle 23. Significant negatively-contributing regions include a cluster across the equator around 2002, and a group of regions in the southern hemisphere towards the end of the cycle, visible as blue patches in all but the bottom-right frame. While the cross-equatorial group is important for reasons discussed above, the majority of regions in the late-emerging cluster might not have had as significant an effect on the current cycle as if they had instead emerged earlier in the cycle, as discussed by \citet{nagy17}, who inserted an extreme active region into a dynamo model simulation at different times throughout a cycle and found that late-emerging regions had the smallest effect. This is because any poleward-advected flux would not have had enough time to reach the pole and cancel with the polar field before the end of the cycle. \bold{The weaker contribution from regions emerging later in the cycle is also evident in Figures \ref{6panelc21}-\ref{6panelc23}, suggesting that it could take at least a few years for regions to reach their asymptotic contributions to the axial dipole moment.} Nevertheless, by analysing Cycles 21 and 23 we see that a lack of disruption from a major cross-equatorial region in Cycle 21 led to a stronger axial dipole moment compared to Cycle 23. The butterfly diagrams again illustrate that the largest contributors are not necessarily the biggest in terms of flux.

\section{Conclusions}\label{conclusions}

Our aim was to test claims that the polar field at the end of Cycle 23 could have been weakened by a small number of large, low-latitude regions. We extracted active region properties from magnetograms using an automated region assimilation technique, and analysed the relationships between these properties and the evolution of the axial dipole moment using a 2D flux transport model. 

We first looked at the effect of keeping regions with the largest final axial dipole moment contribution $D_{\rm rel}\left(t_{\rm end}\right)$ in increments, to see how many were required to obtain a good match with the original axial dipole moment. Using the 500 biggest contributors produced an acceptable axial dipole moment in Cycles 21 and 22, but the lack of small contributions was more damaging in Cycle 23, where at least 750 regions are required to produce an acceptable match. When we only considered the top 10--100 regions, we observed that the odd cycles, especially Cycle 23, struggled to achieve polar field reversal. We attributed this discrepancy to the influence of negatively contributing regions which appear to dominate the axial dipole moment. On the removal of these strongest contributors we found that the axial dipole moment was enhanced, suggesting that the weak polar field at the Cycle 23/24 minimum may have been caused by a small number of extreme regions. When regions were included in order of flux instead of $D_{\rm rel}\left(t_{\rm end}\right)$ there were some differences between cycles, although in each case using the top 80--90\% of the strongest regions was enough to provide a good match to the original axial dipole moment.

We also examined how the final contribution of a single region to the axial dipole moment at the end of the cycle is affected by a region's emergence latitude, flux and initial axial dipole moment, and compared these relationships across Cycle 21, 22 and 23. We found that generally all large contributions to the axial dipole moment emerge below $\pm 20$\textdegree{}, with the largest emerging below $\pm 10$\textdegree{}. This supports the idea that regions emerging at low-latitude can have a large effect on the evolution of the axial dipole moment \citep{cameron13,jiang15}. For our more realistically shaped multipolar regions, we cannot measure the conventional tilt angle, so instead we calculated the more meaningful parameter of initial relative axial dipole moment which takes into account orientation as well as latitude. We found a positive correlation between initial and final $D_{\rm rel}$ within all latitudinal bins in all cycles, but that the constant of proportionality depended on latitude with regions at low latitudes contributing most, whence we concluded that emergence latitude is the dominant parameter controlling the amplification or suppression of the initial dipole moment of a region. This latitude dependence exists because a large dipole moment arises from hemispherical polarity separation, which occurs most effectively when regions emerge tilted and at low latitudes so that cross-equatorial transport of flux can occur \citep{wangsh91,yeates15}. Therefore once we have measured the initial dipole moment of a given region, we can predict its long-term contribution to the dipole moment based purely on its latitude of emergence and the flux decay parameter $\tau$. 

We found that the patterns of regions contributing most to the dipole moment were not consistent across the three cycles. In particular, Cycle 22 contained multiple strong-flux regions which were also some of the largest contributors to the axial dipole moment. This was not the case in Cycles 21 and 23; most large contributors had fluxes of less than \num{2e22}\,Mx, reinforcing that flux alone is not an appropriate measure of contribution. Incidentally, across all cycles there were no significant contributors with fluxes less than \num{1e21}\,Mx, indicating that the smallest regions are not able to drastically alter the axial dipole moment, regardless of emergence latitude. In their coupled surface-interior model, \citet{nagy17} showed that changing BMR tilt and emergence latitude had more immediate consequences than changing flux, unless a very large amount of flux was included. Consequently, if a very large, anti-Joy, anti-Hale region was to emerge close to the equator, it could have a significant detrimental impact on the polar field and hence the amplitude of the next cycle. Following the results of \citet{nagy17} it could even be speculated that, in the most extreme case, such an event could lead to a grand minimum.

As we approach the minimum at the end of Cycle 24, predictions of Cycle 25 will become more reliable, since it becomes less likely that any more large regions which can significantly alter the polar field will emerge. Indeed, from our analysis of the previous three cycles, we only found significant contributors emerging up to the early stages of the descending phase, although that isn't to say such an event is not possible. Indeed, \citet{nagy17} found that `rogue' regions emerging late in the cycle can still have an effect on the following cycle, but this cannot be assessed using our surface flux transport approach, and requires simulation of the interior of the convection zone. For completeness we should go back and repeat this analysis once we reach cycle minimum in a few years' time, using the results to assess any current predictions of Cycle 25.

Some predictions of Cycle 25 have already been made, for example by \citet{hathupton16} and \citet{cameronetal16}, who used two distinct models but came to a similar conclusion: that Cycle 25 will be another weak cycle. However, by incorporating uncertainty in tilt angles and performing multiple simulations, a wider range of cycle amplitudes was found, suggesting that the behaviour of our Sun really does hinge on the random fluctuations in active region properties, highlighting the incurred uncertainty in making \textit{early} forecasts of the next cycle, and that making predictions of future cycles is perhaps futile.

\acknowledgements

TW thanks Durham University Department of Mathematical Sciences for funding his PhD studentship. ARY thanks STFC for financial support. AMJ would like to thank Valent\'in Mart\'inez-Pillet and Scott McIntosh for their support at the National Solar Observatory and High Altitude Observatory. We thank Gordon Petrie for supplying the data in Figure \ref{optbfly}, and magnetogram data were acquired by SOLIS instruments operated by NISP/NSO/AURA/NSF. This work was partly funded by NASA Grand Challenge grant NNH13ZDA001N. We are grateful to the anonymous referee for their useful suggestions.

\appendix

\section{Effect of decay on the axial dipole moment}\label{appendix1}

As mentioned in Section \ref{sect2}, we also remove the decay term from Equation \ref{sfteqn} (i.e. set $\tau\to\infty$) and repeat the optimization and subsequent analysis on the same three cycles. Whilst the equivalent distributions as those shown in the scatterplots of Section \ref{locsize} and butterfly diagrams of Section \ref{loctime} are qualitatively indistinguishable up to a scaling factor, the axial dipole moment profiles for simulations with regions included based on $D_{\rm rel}\left(t_{\rm end}\right)$ as shown in Section \ref{regnum} behave slightly differently, simply because of the lack of decay impacting on cycle minima.

\begin{figure*}
  \resizebox{\hsize}{!}{\includegraphics{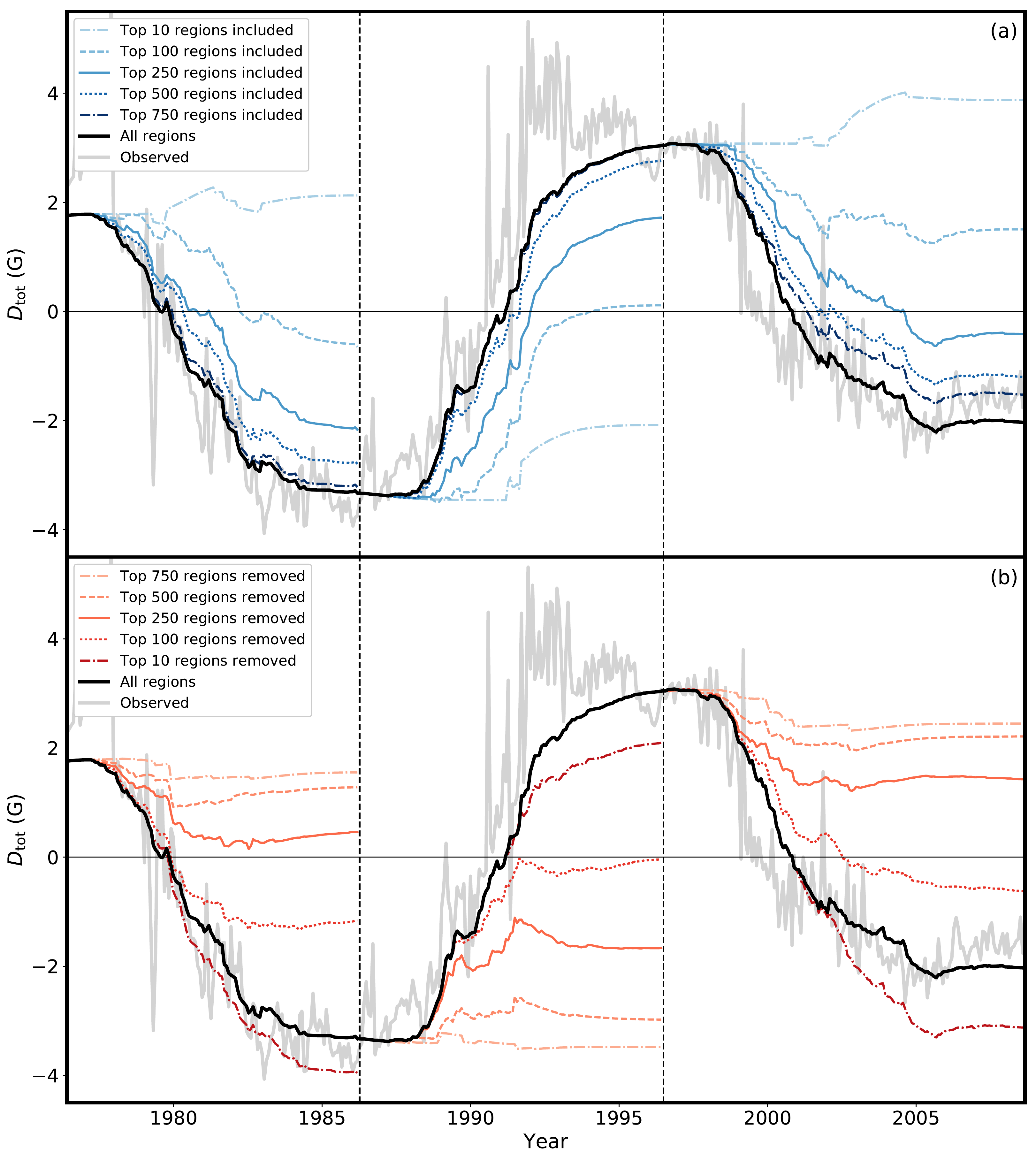}}
  \caption{Evolution of the axial dipole moment for Cycles 21 to 23 with no exponential decay term. Each profile is obtained by: (a) only using a certain number of the biggest contributors to the axial dipole moment, or (b) removing the biggest contributors to the axial dipole moment. Colour intensity is indicative of the number of regions used in each simulation, as shown in the legend. The light grey curve shows the observed axial dipole moment. Vertical dashed lines indicate start/end points of cycles as used in this paper.}
  \label{5profiles_nodecay}
\end{figure*}

The profiles from simulations without decay where only the largest contributors are included are shown in Figure \ref{5profiles_nodecay}(a). With less freedom from fewer parameters, the optimal axial dipole moment does not match the observed counterpart as well when decay is included, but the fit is still acceptable. Again we find that when the top 750 contributors are used, Cycles 21 and 22 are hardly affected but the discrepancy in Cycle 23 is now even more visible than before. When the 100 largest contributors are used, the polar field reverses in Cycles 21 and 22, but not in Cycle 23. Furthermore, polar field reversal is only just achieved with 250 regions, supporting the claim that the biggest contributors from Cycle 23 contribute negatively to the axial dipole moment. For Cycle 21, \citet{wangsh91} found that about 54\% of the axial dipole moment came from about 10.7\% of regions, and here we find a similar result (blue curve). In fact, we find the same outcome for Cycle 22 but not for Cycle 23.

Figure \ref{5profiles_nodecay}(b) shows the axial dipole moment evolution when the strongest regions are removed from each cycle. With no exponential decay, the deficit created by the removal of the top 10 regions of Cycle 23 is even clearer here than in Figure \ref{5profiles}(b), highlighting the detrimental effect of those contributors with negative $D_{\rm rel}\left(t_{\rm end}\right)$.

\bibliographystyle{aasjournal}
\bibliography{mybiblio}

\end{document}